\documentclass[english,12pt,aps,prd,a4paper,preprintnumbers, floatfix,nofootinbib,showpacs,superscriptaddress, notitlepage]{revtex4-1}

\usepackage{pslatex}
\usepackage[pdftex]{graphicx}
\usepackage{psfrag}
\usepackage{epsfig}
\usepackage[usenames,dvipsnames]{color}
\usepackage{cancel}
\usepackage{slashed}
\usepackage{amssymb}
\usepackage{amsmath}
\usepackage{enumerate}
\usepackage{multirow}
\usepackage{natbib}
\usepackage[normalem]{ulem}
\usepackage[colorlinks=true,citecolor=darkred,urlcolor=darkred,pdfborder={0 0 0}]{hyperref}

\definecolor{darkred}{rgb}{0.6,0,0}
\definecolor{dbrown}{rgb}{0.4,0.26,0.13}
\definecolor{linkcolor}{rgb}{0,0,0.5}
\definecolor{dgreen}{rgb}{0.,0.6,0.}

\begin{document}
\title{Neutrino magnetic and electric dipole moments: From
  measurements to parameter space}%
\author{D. Aristizabal Sierra}%
\email{daristizabal@ulg.ac.be}%
\affiliation{Universidad T\'ecnica
  Federico Santa Mar\'{i}a - Departamento de F\'{i}sica\\
  Casilla 110-V, Avda. Espa\~na 1680, Valpara\'{i}so, Chile}%
\author{O. G. Miranda}%
\email{omr@fis.cinvestav.mx}%
\affiliation{Departamento de F\'{i}sica, Centro de Investigaci\'on y
  de Estudios Avanzados del IPN, Apartado Postal 14-740 07000 Mexico,
  Distrito Federal, Mexico}%
\author{D. K. Papoulias}%
\email{d.papoulias@uoi.gr}%
\affiliation{Department of Physics, University of Ioannina GR-45110
  Ioannina, Greece}%
\author{G. Sanchez Garcia}%
\email{gsanchez@fis.cinvestav.mx}%
\affiliation{Departamento de F\'{i}sica, Centro de Investigaci\'on y
  de Estudios Avanzados del IPN, Apartado Postal 14-740 07000 Mexico,
  Distrito Federal, Mexico}%
\begin{abstract}
  Searches for neutrino magnetic moments/transitions in low energy
  neutrino scattering experiments are sensitive to effective couplings
  which are an intricate function of the Hamiltonian parameters. We
  study the parameter space dependence of these couplings in the
  Majorana (transitions) and Dirac (moments) cases,  as well as the impact of
  the current most stringent experimental upper limits on the
  fundamental parameters. In the Majorana case we find that for
  reactor, short-baseline and solar neutrinos, CP violation can be
  understood as a measurement of parameter space vectors
  misalignments. The presence of nonvanishing CP phases opens a blind
  spot region where---regardless of how large the parameters are---no
  signal can be observed in either reactor or short-baseline
  experiments. Identification of these regions requires a combination of
  different data sets and allows for the determination of those CP
  phases. We point out that stringent bounds not necessarily imply
  suppressed Hamiltonian couplings, thus allowing for regions where
  disparate upper limits can be simultaneously satisfied. In contrast,
  in the Dirac case stringent experimental upper limits necessarily
  translate into tight bounds on the fundamental couplings. In terms
  of parameter space vectors, we provide a straightforward mapping of experimental information
  into parameter space.
\end{abstract}
\maketitle
%\tableofcontents
% -------------------
% Sec: Introdcution
% -------------------
\section{Introduction}
\label{sec:intro}
For decades, experimental signals of magnetic and electric dipole moments
have been searched for in a large variety of environments (for a review see Ref.~\cite{Giunti:2014ixa}). Their
distinctive feature---regardless of neutrino flavor---is that of
spectral distortions at low recoil energies, thus making detectors
with low recoil energy resolutions an ideal tool for such
searches. Rather than being controlled by a single parameter, the size
of the distortion is instead governed by an effective coupling,
$\mu_{\nu_\alpha}^\text{eff}$, with flavor dependence $\alpha$
determined by the incoming neutrino flux. In the absence of such
signal and using neutrino-electron elastic scattering as well as
coherent elastic neutrino-nucleus scattering (CE$\nu$NS),
short-baseline accelerator facilities have placed competitive bounds on
the electron and muon effective couplings
\cite{LSND:2001akn,DONUT:2001zvi,Kosmas:2015sqa,Miranda:2019wdy}. Solar
neutrino experiments, in particular BOREXINO, have placed even more
stringent limits on the electron effective coupling, which along with
the GEMMA reactor experiment and XENON1T set the most tight bounds on this
coupling at the laboratory level \cite{Borexino:2017fbd,Beda:2012zz,XENON:2020rca}.

Forthcoming low threshold reactor experiments
\cite{Hakenmuller:2019ecb,Bonet:2020awv,CONNIE:2019swq,Strauss:2017cuu,Agnolet:2016zir,Choi2020NeutrinoEO,RED-100:2019rpf,Billard:2016giu,Wong:2015kgl,SBC:2021yal},
short-baseline neutrino detectors \cite{Baxter:2019mcx}, multi-ton
dark matter (DM) direct detection experiments
\cite{Aalbers:2016jon,Aprile:2015uzo,LUX-ZEPLIN:2018poe} and possibly
even decay-in-flight neutrino facilities
\cite{AristizabalSierra:2021uob} will further scrutinize these
regions. Indeed, searches for neutrino magnetic couplings along with
searches for other novel interactions in the neutrino sector including
nonstandard interactions (NSI), neutrino generalized interactions
(NGI) as well as long range scalar and vector interactions, are among
the experimental phenomenological targets of these type of experiments
(see
e.g. \cite{AristizabalSierra:2020zod,Shoemaker:2017lzs,Shoemaker:2021hvm,Denton:2018xmq,AristizabalSierra:2019ykk,AristizabalSierra:2019ufd,Majumdar:2021vdw,Cadeddu:2020nbr,Dutta:2017nht,AristizabalSierra:2017joc}). As
pointed out recently, the presence of these interactions may actually
have deep implications for DM direct detection searches
\cite{AristizabalSierra:2021kht}.

Motivated by these experiments, in this paper we study the
implications of measurements (or upper bounds, resulting from the
absence of a signal) of neutrino magnetic and electric dipole
moments/transitions~\cite{Fujikawa:1980yx,Schechter:1981hw,Pal:1981rm,Kayser:1982br,Nieves:1981zt,Shrock:1982sc}
on parameter space. Focusing on both the Majorana and full diagonal
Dirac cases we: (i) Study possible basis invariants that enable a
basis-independent mapping of experimental information into parameter
space, (ii) derive general matricial expressions for the effective
couplings in short-baseline, reactor and solar neutrino experiments
(in the mass and flavor eigenstate bases), (iii) introduce a parameter
space vector notation that allows a straightforward mapping of data
into parameter space and allows as well the identification of the role
played by CP phases in opening blind spots in the Majorana
case\footnote{Here with blind spots we mean regions in parameter space
  where an effective coupling vanishes despite the fundamental
  parameters being different from zero. For a given experiment this
  translates in the absence of a signal.}, (iv) perform a general
mapping of LSND \cite{LSND:2001akn}, GEMMA \cite{Beda:2012zz} and
XENON1T \cite{XENON:2020rca} experimental data into parameter space,
that enables demonstrating the viability of reconciling the LSND bound
with the more stringent GEMMA limit.

The remainder of this paper is organized as follows. In
Sec. \ref{sec:NMEDM} we introduce the couplings involved in our analysis, along with the notation employed throughout the paper. In
Sec. \ref{sec:scattering} we present a general analysis, valid for any
value of the Dirac CP phase, of the effective couplings in
short-baseline and reactor experiments, as well as in solar neutrino
experiments, in both mass and flavor eigenstate bases. In
Sec. \ref{sec:majorana_case}, we specialize the discussion to the
Majorana case and introduce a vector notation that allows for a
straightforward mapping of experimental data to parameter space. In
Sec. \ref{sec:Dirac_case}, we discuss neutrino magnetic moments in the
Dirac case, while in Sec. \ref{sec:conclusions} we summarize and
present our conclusions. In Appendices \ref{sec:average_solar} and
\ref{sec:majorana_case_leton_mixing_matrix} we calculate the phase
average for solar neutrinos and demonstrate the connection between
blind spots and the unitarity of the lepton mixing matrix for the
Majorana case.
% -----------
% Section-I
% -----------
\section{Neutrino magnetic and electric dipole moment interactions}
\label{sec:NMEDM}
The structure of the interactions we are interested in, depends on
whether neutrinos are Dirac or Majorana particles. For the Dirac case---in the
flavor eigenstate basis---they are described by the effective
Lagrangian~\cite{Grimus:2000tq}
\begin{equation}
  \label{eq:Lagrangian_D}
  \mathcal{H}_D = -\frac{1}{2}\overline{\nu}\sigma_{\mu\nu}
  \left(\mu + i \epsilon\gamma_5\right)
  \nu\,F^{\mu\nu}=-\frac{1}{2}\overline{\nu_R}\sigma_{\mu\nu}\,\lambda\,\nu_L\,F^{\mu\nu}
  + \text{H.c.}\ ,
\end{equation}
where $\nu^T=(\nu_e,\nu_\mu,\nu_\tau,\nu_s\cdots)$ with $\nu_s$ denoting possible
 sterile neutrino states, and $\nu=\nu_L + \nu_R$ being the Dirac
neutrino fields. The matrices $\mu$ and $\epsilon$ are $N\times N$
Hermitian matrices in flavor space that satisfy $\mu^\dagger =\mu$ and
$\epsilon^\dagger=\epsilon$, as required by the Hermiticity of the
Hamiltonian. Their entries correspond to neutrino magnetic and
electric dipole moments ($i=j$) and transitions ($i\neq j$), which
follow from the zero momentum transfer limit of the magnetic and
electric dipole form factors, $F_M(q^2)$ and $F_E(q^2)$~\cite{Giunti:2014ixa}. The following
combination of $\mu$ and $\epsilon$ matrices defines
\begin{equation}
  \label{eq:lambda_matrix}
  \lambda = \mu - i\epsilon\ ,
\end{equation}
a matrix that proves to be rather useful when describing neutrino
scattering induced by these interactions. In the Majorana case the
effective Hamiltonian, written as well in the flavor eigenstate basis
reads \cite{Schechter:1981hw}
\begin{equation}
  \label{eq:Lagrangian_M}
  \mathcal{H}_M = -\frac{1}{4}\nu^TC^{-1}\sigma_{\mu\nu}
  \left(\mu + i\epsilon\gamma_5\right)
  \nu\,F^{\mu\nu}=-\frac{1}{4}\nu_L^TC^{-1}\sigma_{\mu\nu}
  \,\lambda\,\nu_L\,F^{\mu\nu} + \text{H.c.}\ .
\end{equation}
Here the massive Majorana fields are given by $\nu=\nu_L+\nu_L^C$ and
the matrices $\mu$ and $\epsilon$, in addition to being Hermitian, are
as well complex antisymmetric: $\mu^T=-\mu$ and
$\epsilon^T=-\epsilon$. An implication of their antisymmetric
character is that in the Majorana case the neutrino and electric
dipole moments vanish and transitions are purely imaginary numbers.

Relations between the flavor and mass eigenstate bases follow from the
diagonalization of the neutrino mass matrices ($m_D$ for Dirac and
$m_M$ for Majorana). In the Dirac case it proceeds as in the charged
fermion sector through bi-unitary transformations. In the Majorana
case, instead, the symmetric character of the neutrino mass matrix
enables diagonalization solely through the lepton mixing
matrix. Quantitatively one has
\begin{equation}
  \label{eq:Dirac_basis}
  \nu_L=U\cdot \nu_L^\prime
  \quad\text{and}\quad
  \nu_R=V\cdot\nu_R^\prime
  \quad \Rightarrow\quad
  V^\dagger\cdot m_D\cdot U = \hat m_D
  \quad\text{and}\quad
  U^T\cdot m_M\cdot U=\hat m_M\ .
\end{equation}
Here primed fields refer to mass eigenstates. Note that the unitary
matrix $V$ enters only in the Dirac mass term, and has no effect in
any other coupling nor in neutral or charged currents. This means that
it has no physical character and so any choice for its parametrization
should be as good as any other. The most ``extreme'' choice would be
rotating it away, which is equivalent to $V=\mathbb{I}$. If one does
so, the diagonalization in the Dirac case in
Eq. (\ref{eq:Dirac_basis}) reduces to $m_D\cdot U = \hat m_D$, from
which one can see that the Dirac mass matrix
$m_D=\hat m_D\cdot U^\dagger$ is constrained. It involves one
CP-violating phase and six independent real parameters, rather than
six CP phases and nine real parameters as it should. Following this
argument we then fix $V=U^*$, a choice that enables the relation for
the transformation from the flavor to the mass eigenstate basis for
$\lambda$ to have the same structure in the Dirac and Majorana cases
(see discussion below).

\begin{table}
  \renewcommand{\tabcolsep}{0.1cm}
  \renewcommand{\arraystretch}{1.2}
  \centering
  \begin{tabular}{|c|c|c|c||c|c|c|c|}\hline
    \multicolumn{4}{|c||}{\textbf{Dirac}} & \multicolumn{4}{|c|}{\textbf{Majorana}}\\\hline\hline
    Matrix & Type & Moduli & CP phases & Matrix & Type & Moduli & CP phases
    \\\hline
    $\lambda$ & $\mathbb{C}$ & $N^2$ & $N(N-1)$ & $\lambda$ & $\mathbb{C}$ antisymmetric & $N(N-1)/2$ & $N(N-1)/2-1$
    \\\hline
    $\mu$ & Hermitian & $N(N+1)/2$ & $N(N+1)/2$ & $\mu$ & $\mathbb{C}$ antisymmetric  & $N(N-1)/2$ & $N(N-1)/2$
    \\\hline
    $\epsilon$ & Hermitian & $N(N+1)/2$ & $N(N+1)/2$ & $\epsilon$ & $\mathbb{C}$ antisymmetric & $N(N-1)/2$ 
    & $N(N-1)/2$
    \\\hline
  \end{tabular}
  \caption{Number of physical parameters, including CP-violating phases, that define
    the neutrino magnetic and electric dipole matrices as well as the
    $\lambda=\mu - i\epsilon$ matrix. Results are shown for a general 
    $N\times N$ ($N$ active and $N$ sterile) neutrino scenario in both the Dirac and
    Majorana neutrino cases.}
  \label{tab:num_parameters}
\end{table}

The coupling matrices in Eqs.~(\ref{eq:Lagrangian_D}) and
(\ref{eq:Lagrangian_M}) involve a particular number of physical
CP-violating phases. Its counting is as follows. In the Dirac case,
the magnetic and electric dipole $N\times N$ matrices---being
Hermitian---involve $N(N+1)/2$ moduli and $N(N-1)/2$ CP-violating
phases. Since only diagonal phases can be removed by phase rotations
of the neutrino fields, all the phases are physical. The general
Majorana case, instead, involves $N(N-1)/2$ moduli and the same number
of physical CP-violating phases. Despite the properties of Hermiticity
of the magnetic and electric dipole matrices, in the Dirac case
$\lambda$ is a complex $N\times N$ matrix. Accordingly, it involves
$2N^2$ parameters, $N^2$ moduli and $N^2$ phases, of which $N$ can be
removed by phase rotations of the neutrino fields, resulting in
$N(N-1)$ physical phases.  In the Majorana case the combination keeps
being antisymmetric, but one CP phase can be removed by a single field
redefinition. In the Majorana case then $\lambda$ involves $N(N-1)/2$
moduli and $N(N-1)/2-1$ physical phases. A summary of the physical
parameters, including the CP-violating phases that define each matrix
in each case, is shown in Tab.~\ref{tab:num_parameters}. Taking into
account this parameter counting, the coupling matrix (in the
$3\times 3$ case) for Dirac and Majorana neutrinos in the mass
eigenstate basis can be parametrized as follows:

\begin{equation}
  \label{eq:lambda_3times3_parametrization}
  \lambda_D^\prime=
  \begin{pmatrix}
    \lambda_{11}e^{i\varphi_{11}} & |\lambda_{12}|e^{i\varphi_{12}} & |\lambda_{13}|e^{i\varphi_{13}}\\
    |\lambda_{21}|e^{i\varphi_{21}} & \lambda_{22}e^{i\varphi_{22}} & |\lambda_{23}|e^{i\varphi_{23}}\\
    |\lambda_{31}|e^{i\varphi_{31}} & |\lambda_{32}|e^{i\varphi_{32}} & \lambda_{33}e^{i\varphi_{33}}\\
  \end{pmatrix}\ ,
  \qquad
  \lambda_M^\prime=
  \begin{pmatrix}
     0 & |\Lambda_3|e^{i\varphi_3} & -|\Lambda_2|\\
     -|\Lambda_3|e^{i\varphi_3} & 0 & |\Lambda_1|e^{i\varphi_1}\\
     |\Lambda_2| & -|\Lambda_1|e^{i\varphi_1} & 0 \\
  \end{pmatrix}\ ,
\end{equation}
where the notation $\Lambda_i=\epsilon_{ijk}\lambda_{jk}$ has been
adopted in the Majorana case\footnote{For the Majorana phases a
  different notation has been previously used in
  Refs. \cite{Canas:2015yoa,Miranda:2019wdy}. The translation is
  straightforward: $\varphi_1=\xi_3=\zeta_1 -\zeta_2$ and
  $\varphi_3=\xi_1=\zeta_3-\zeta_2$.}. With the aid of the
transformation bases in Eqs. (\ref{eq:Dirac_basis}), bearing in mind
that we have chosen $V=U^*$, the relations between the coupling matrix
in the mass and flavor bases read \footnote{We will see that the choice of $V$ does not affect the form of the effective magnetic moment.}:
\begin{equation}
  \label{eq:lambda_flavor_basis}
  \lambda_D^\prime = U^T\cdot \lambda_D\cdot U\ ,\qquad
  \lambda_M^\prime = U^T\cdot \lambda_M\cdot U\ .
\end{equation}
One can see that if rather $V$ is rotated away the relation between
the flavor and mass eigenstate bases in the Dirac case reduces to
$\widetilde{\lambda}_D^\prime =\lambda_D\cdot U$.  This, however, does
not mean that results involving scattering processes are affected by
this choice, something expected due to the nonphysical character of
$V$ (see discussion in Sec. \ref{sec:scattering}).
% --------------
% Section
% --------------
\section{Scattering processes induced by magnetic and electric dipole
  moments}
\label{sec:scattering}
Weak processes produce neutrinos
  as flavor eigenstates.  As they propagate towards a detector located
at a distance $L$, they may be subject to flavor oscillations,
depending on $L$ and on their average energy. As they reach the
detector, the interactions in Eqs. (\ref{eq:Lagrangian_D}) and
(\ref{eq:Lagrangian_M}) induce (electromagnetic (EM))
scatterings with the target material. Depending on the neutrino energy
 spectrum, the scatterers can involve
electrons, nuclei, nucleons, or \sout{even} quarks. For neutrino
energies relevant for CE$\nu$NS ($E_\nu\lesssim 100\,$MeV),  as those coming 
from reactors, or pion decay-at-rest ($\pi$-DAR) neutrino
sources, scattering is dominated by neutrino-nucleus and
neutrino-electron processes. For energies as those in decay-in-flight
neutrino beams,  e.g., NuMI or DUNE, the EM scattering
is---instead---driven by neutrino-nucleon processes or quarks.

EM $t$-channel scattering processes
$\nu_\alpha + X \rightarrow \nu_\alpha + X$ ($X=N,e,n,q$) involve
chirality flip, in contrast to those induced by electroweak (EW)
interactions. Thus, they do not interfere with the SM contribution. In
the presence of the new EM interactions the differential scattering
cross section is then a sum of two terms:
$d\sigma/dE_r = d\sigma_\text{EW}/dE_r + d\sigma_\text{EM}/dE_r$,
where $d\sigma_\text{EW}/dE_r$ and $d\sigma_\text{EM}/dE_r$ refer to
the SM and EM components, respectively. The second contribution, being of EM
origin, exhibits a Coulomb divergence which becomes particularly
pronounced at small momentum transfer, $q$. For this reason is that
the most suitable targets for EM neutrino interaction searches are
neutrino-electron elastic scattering and CE$\nu$NS. Experiments with
sensitivities to these processes operate at rather low thresholds and
so are able---in principle---to identify possible spectral distortions
generated by the new interactions. In contrast, experimental setups
involving neutrino-nucleon or neutrino-quark interactions operate at
higher thresholds. Identification of spectral distortions require then
$\lambda$ couplings whose values have been already ruled out, implying
that decay-in-flight neutrino beams are not suitable for neutrino EM
interaction searches. In what follows we then focus our discussion on
neutrino-electron elastic scattering and CE$\nu$NS.

The EM neutrino-electron elastic scattering differential cross section
was first calculated by Vogel and Engel in Ref. \cite{Vogel:1989iv},
without taking into account flavor oscillation effects. However, for a
neutrino flavor (weak) eigenstate $\nu_\alpha$ ($\alpha=e,\mu,\tau$)
produced at $t^\prime=0$ and further detected at $L$ at a time
$t^\prime=t$ these effects are unavoidable.  References
\cite{Grimus:1997aa,Beacom:1999wx} have taken them into account by
considering propagation in the mass eigenstate basis and
neutrino-electron elastic scattering. Doing so, they have found the
following differential cross section:
\begin{equation}
  \label{eq:EM_neutrino_electron_scattering_xsec}
  \frac{d\sigma_\text{EM}}{dE_r}=\frac{\pi\alpha^2}{m_e^2}
  \left(\frac{1}{E_r} - \frac{1}{E_\nu}\right)
  \frac{\overline{\mu}_{\nu_\alpha}^2(L,E_\nu)}{\mu_B^2}\ ,
\end{equation}
where the dimensionful coupling
$\overline{\mu}_{\nu_\alpha}(L,E_\nu) =  {\mu}_{\nu_\alpha}(L,E_\nu) \mu_B $ has been normalized to the Bohr
magneton $\mu_B=1/2/m_e$.  After rescaling the previous equation by
$Z^2 F_W^2(q^2)$, with $F_W(q^2)$ being the
  nuclear form factor, this result applies as well to
CE$\nu$NS~\cite{Papoulias:2015iga}.

In the context of an ``agnostic'' phenomenological analysis,
Eq. (\ref{eq:EM_neutrino_electron_scattering_xsec}) can be used
without specifying the parameter space coupling function
$\mu_{\nu_\alpha}^2\equiv \langle \mu_{\nu_\alpha}^2(L,E_\nu)
\rangle$\footnote{For
  the cases in which the effective coupling has a $L/E_\nu$ dependence
  an average over this variable has to be done, in the same vein one
  does with oscillation probabilities (see
  e.g. \cite{Giunti:2007ry}). A normalized Gaussian function can be
  used for that aim.}.  Such an approach
allows placing constraints on $\mu_{\nu_\alpha}^2$, which can then be
mapped into the $\lambda_{ij}$ parameter space~\cite{Canas:2015yoa,Miranda:2019wdy,Miranda:2020kwy,Miranda:2021kre}. However, this second step is rarely done, and results obtained for $\mu_{\nu_\alpha}^2$
are directly presented instead.  This can lead to
misleading interpretations of results derived from different data
sets. It can potentially imply that a limit derived from, say, reactor
data is taken to be universal regardless of the experimental context
to which one is comparing to.  To avoid confusion, and to make
sure that a certain limit is properly applied, the mapping into the $\lambda_{ij}$ parameter space should
then be understood as mandatory.

To begin with, we then write $\mu_{\nu_\alpha}^2(L,E_\nu)$ for incoming
neutrinos in the most general way in terms of the couplings in the
mass eigenstate basis \cite{Grimus:1997aa,Beacom:1999wx}
\begin{equation}
  \label{eq:mu_nu_general}
  \mu_{\nu_\alpha}^2(L,E_\nu)=\sum_{j=1}^3\left|U^*
    \cdot P_j(L,E_\nu) \cdot \lambda^{\prime T}\right|^2\ ,
\end{equation}
where, in contrast to
Refs. \cite{Grimus:1997aa,Beacom:1999wx,Giunti:2014ixa}, we have
employed matrix notation since this allows for an easier
identification of basis transformation effects.  The leptonic mixing
matrix in this expression accounts for the fact that flavor
eigenstates are a superposition of mass eigenstates.  Thus,  depending
on the environment at which neutrinos have been produced and further
propagated,  this matrix can be either that in vacuum or in matter.  The
$3\times 3$ matrices $P_j$ in Eq. (\ref{eq:mu_nu_general}) are
diagonal matrices defined as $P_j=\text{diag}(P_{1j},P_{2j},P_{3j})$,
where $P_{ij}=e^{-i\Delta_{ij}}$ and the phases are given by
$\Delta_{ij}=\Delta m_{ij}^2L/2/E_\nu$. Since neutrino oscillation data
provide information on $\Delta m_{21}^2$ and on $|\Delta m_{31}^2|$
\cite{deSalas:2020pgw}, the $P_{ij}$ entries should be expressed in
terms of just two phases $\Delta_{21}$ and $\Delta_{31}$. Taking into
account that $\Delta m_{21}^2\ll |\Delta m_{31}^2|$ one finds for the
\textit{normal spectrum case} $P_{21}=e^{-i\Delta_{21}}$,
$P_{31}=e^{-i\Delta_{31}}$ and the following relations: $P_{ii}=1$,
$P_{12}=P_{21}^*$, $P_{13}=P_{23}=P_{31}^*$ and $P_{32}=P_{31}$. For
the \textit{inverted spectrum case}, instead, $P_{31}=e^{i\Delta_{31}}$,
$P_{13}=P_{32}=P_{31}^*$ and $P_{23}=P_{31}$ with all the other
relations as in the normal case.  By shifting
$\Delta_{ij}\to -\Delta_{ij}$, Eq. (\ref{eq:mu_nu_general}) is valid for
antineutrinos as well \cite{Giunti:2014ixa}.

From Eq. (\ref{eq:mu_nu_general}) it is clear that different
experimental setups imply different effective coupling functions. There are two
variables which are key to this statement: (i) the oscillation phases
$\Delta_{21}$ and $\Delta_{31}$, and (ii) the incoming neutrino flavor state
index $\alpha$. These two variables are independent of the basis one
chooses for the description of the scattering process.  Focusing first
on (i), we then write the oscillation phases according to
\begin{align}
  \label{eq:oscillation_phases_solar}
  \Delta_{21}=\frac{\Delta m_{21}^2}{2E_\nu}L &= 3.8\times 10^{-4} 
  \left(\frac{\Delta m_{21}^2}{7.50\times 10^{-5}\,\text{eV}^2}\right)
  \left(\frac{10\,\text{MeV}}{2E_\nu}\right)
  \left(\frac{L}{10\,\text{m}}\right)\ ,
  \\
  \label{eq:oscillation_phases_atm}
  \Delta_{31}=\frac{\Delta m_{31}^2}{2E_\nu}L &= 1.3\times 10^{-2} 
  \left(\frac{\Delta m_{31}^2}{2.55\times 10^{-3}\,\text{eV}^2}\right)
  \left(\frac{10\,\text{MeV}}{2E_\nu}\right)
  \left(\frac{L}{10\,\text{m}}\right)\ ,
\end{align}
where best-fit point values for $\Delta m_{21}^2$ and
$\Delta m_{31}^2$ (normal ordering case) have been used
\cite{deSalas:2020pgw} along with typical values for $L$ and $E_\nu$.
For the processes we are interested in,  the experimental environments that
matter are short-baseline reactors, $\pi$-DAR neutrino facilities,
and solar neutrino detectors. For reactors and $\pi$-DAR neutrino
sources, the numbers quoted in
Eqs. (\ref{eq:oscillation_phases_solar}) and
(\ref{eq:oscillation_phases_atm}) apply, and one can fairly assume
$\Delta_{ij}\to 0$ and so $P_j\to \text{diag}(1,1,1)$ for all
$j$. Moreover, since matter effects are absent in these cases, $U$
matches the leptonic mixing matrix in
vacuum.  Equation (\ref{eq:mu_nu_general}) specialized to these two cases
can then be written according to:
\begin{equation}
  \label{eq:reactor_decat_at_rest_mu_nu}
  \text{Reactor and $\pi$-DAR (mass basis):}\quad\mu_{\nu_\alpha,}^2(L\to0)=\sum_{j=1}^3\left|\left(U^*
      \cdot \lambda^{\prime T}\right)_{\alpha j}\right|^2
  = \left(U\cdot \lambda^{\prime \dagger}\cdot \lambda^\prime
    \cdot U^\dagger\right)_{\alpha\alpha}\ .
\end{equation}
From now on, we will use the notation $\mu_{\nu_\alpha,}^2$ to denote this no-oscillation case. 
Notice that as a consequence, in these cases the expression of the effective coupling for neutrinos and anti-neutrinos is the same.
For solar neutrinos the discussion is more subtle. Interference terms
in Eq. (\ref{eq:mu_nu_general}) involve the propagation phases
$\Delta_{21}$ and $\Delta_{31}$, which cannot be dropped as in the
previous cases ($L$ is fixed by the Sun-Earth distance and so
$L/E_{\nu}\gg 1$). In this case then $\mu^2_\nu(L,E_\nu)$ should be averaged
over $L/E_\nu$. As already mentioned, this can be done by assuming the
smearing function to be a normalized Gaussian function with median
$\mu=\langle L/E_\nu \rangle$ and variance $\sigma$. Doing so one gets
for the average values of the propagation phases the following result
(for completeness, we present the details in Appendix
\ref{sec:average_solar})
\begin{equation}
  \label{eq:phases_averaged}
  \langle e^{\pm \Delta m_{ij}^2 L/2/E_\nu}\rangle = e^{\pm \Delta m_{ij}^2 L/2/E_\nu}
  e^{-\Delta m_{ij}^4 \langle L/E\rangle^2/8}\ ,
\end{equation}
which implies that interference terms in the solar case are (strongly)
exponentially suppressed and so can be ignored. With this simplification,
Eq. (\ref{eq:mu_nu_general}) thus reads:
\begin{equation}
  \label{eq:mu_nu_solar}
  \text{Solar neutrinos (mass basis):}\quad\langle \mu_{\nu, \text{sol}}^2\rangle\equiv \mu_{\nu, \text{sol}}^2=
  \sum_{k=1}^3\left|U_{\alpha k}^\text{M}\right|^2
    \left(\lambda^{\prime \dagger}\cdot \lambda^\prime\right)_{kk}\ ,
\end{equation}
where the leptonic mixing matrix $U^M$ involves mixing angles
in matter which implies that the coupling is still energy dependent.

A relevant point---already mentioned at the end of
Sec. \ref{sec:NMEDM}---has to do with the impact of the change of
basis in Eqs. (\ref{eq:reactor_decat_at_rest_mu_nu}) and
(\ref{eq:mu_nu_solar}). Or, in other words, on whether $\mu_\nu^2$ as
defined by these equations is basis independent or not. This question
is actually crucial for two reasons. First of all because the
nonphysical character of $V$, enables parametrizing it
arbitrarily. Secondly, because it tells us how robust results derived
in a particular basis are. Let us first consider the Dirac case and
the most general transformation between the mass and flavor eigenstate
bases: $\lambda^\prime = V\cdot \lambda\cdot U$. One can see that
Eqs. (\ref{eq:reactor_decat_at_rest_mu_nu}) and (\ref{eq:mu_nu_solar})
become
\begin{alignat}{2}
  \label{eq:flavor_reactor_DAR}
  \text{Reactor and $\pi$-DAR (flavor basis):}\quad&
  \mu_{\nu_\alpha}^2 = \left(\lambda^\dagger\cdot \lambda\right)_{\alpha\alpha}\ ,
  \\
  \label{eq:flavor_solar}
  \text{Solar neutrinos (flavor basis):}\quad&
  \mu_{\nu_\alpha}^2 = \sum_{k=1}^3|U_{\alpha k}^\text{M}|^2
  \left(U^\dagger\cdot \lambda^\dagger\cdot \lambda\cdot U\right)_{kk}\ .
\end{alignat}
These results do not depend on $V$ and therefore imply  that any choice for
this matrix is as good as any another. They are valid as well in the
Majorana case, thus demonstrating that the flavor effective neutrino
magnetic moment coupling is not basis independent. Interesting,
however, is the fact that for the reactor and $\pi$-DAR cases, the sum of
these quantities over flavor is a basis invariant
\begin{equation}
  \label{eq:basis_invariant}
  \sum_{\alpha=e,\mu,\tau}\mu_{\nu_\alpha}^2
  = \text{Tr}\left(\lambda^{\prime\dagger}\cdot \lambda^\prime\right)
  = \text{Tr}\left(\lambda^\dagger\cdot \lambda\right)\ .
\end{equation}
This implies that if one can access data sets which involve the three
flavors (e.g. GEMMA, COHERENT and DONUT), the mapping from these
measurements to the fundamental parameters can be done in a
basis-independent way. One can also see that in the mass eigenstate
basis, CP-violating effects are present in the reactor and $\pi$-DAR
effective coupling (Eq. (\ref{eq:reactor_decat_at_rest_mu_nu})), while they are
absent in the solar effective coupling
(see Eq. (\ref{eq:mu_nu_solar})). Conversely, in the flavor eigenstate
basis they play a role in the solar coupling, while they have no effect in the reactor
and $\pi$-DAR effective parameter.  This is somehow expected given the physical nature
of the phases.  A basis choice hides these phases from a certain effective
coupling but at the same time exposes them in the other. In an analysis
including reactor, $\pi$-DAR, and solar data, one chooses a basis and thus if
the physics responsible for electric and magnetic dipole moments is
CP-violating, then the effects of the CP-violating phases will necessarily
show up (provided the phases are large enough).

Regarding (ii), for the reactor case we have $\alpha=e$
($\overline{\nu}_e$), while for $\pi$-DAR neutrinos $\alpha=\mu$ and
$\alpha=e$ for the prompt ($\nu_\mu$) and delayed
($\overline{\nu}_\mu, \nu_e$) incoming neutrino states,
respectively. For solar $\alpha=e$ ($\nu_e$). In what follows, using
Eqs. (\ref{eq:reactor_decat_at_rest_mu_nu}), (\ref{eq:mu_nu_solar}),
(\ref{eq:flavor_reactor_DAR}) and (\ref{eq:flavor_solar}), we then
detemine the impact of current more stringent bounds on the
fundamental parameters for the Majorana and Dirac neutrino cases
according to Eq. (\ref{eq:lambda_3times3_parametrization}),
differentiating beween the CP-conserving and CP-violating scenarios.

% ----------------
% Section
% ----------------
\section{Majorana neutrino transition magnetic and electric dipole
  moments}
\label{sec:majorana_case}
% ----------------
% Section
% ----------------

\subsection{Effective couplings in reactor and short baseline accelerator experiments}
\label{sec:effective_couplings_Majorana}
We begin with the Majorana case for short distance sources in the mass
basis. To simplify the notation, we see that
Eq. (\ref{eq:reactor_decat_at_rest_mu_nu}), combined with the Majorana
coupling matrix in Eq. (\ref{eq:lambda_3times3_parametrization}) can
be recast in terms of parameter space vectors, namely:
\begin{equation}
  \label{eq:coupling_in_terms_of_vectors}
  \mu_{\nu_\alpha}^2
  = \sum_{i=1}^3\left|\Lambda_i\right|^2
  - \left|\sum_{i=1}^3\vec{\Lambda}_{i\alpha}\right|^2
  = \left|\vec{\Lambda}\right|^2
  - \left|\sum_{i=1}^3\vec{\Lambda}_{i\alpha}\right|^2\ .
\end{equation}
We denote $\vec{\Lambda}= \Lambda_i \widehat{e}_i$, and we define the
flavor-dependent vectors, which are in general not orthogonal, as:
\begin{equation}
  \label{eq:vector_definition}
  \vec{\Lambda}_{i\alpha} = \left|U_{\alpha i}\right|
  \left|\Lambda_i\right|\widehat{e}_{i\alpha}\ ,
  %f_{i\alpha}\;\Lambda_i\;\widehat{e}_{i\alpha}\ ,
\end{equation}
with $\widehat{e}_{i\alpha}$ being unit vectors. Within this
formalism, the misalignment between these vectors is determined by the
phases of the magnetic moment couplings, and can be interpreted as a
measure of the amount of CP violation the new physics comes along
with. They are weighted according to the entries of the lepton mixing
matrix $U_{\alpha i}$, which together with the misalignment products,
$\widehat{e}_{i\alpha}\cdot \widehat{e}_{j\alpha}$, determine their
size (for details see Appendix
\ref{sec:majorana_case_leton_mixing_matrix}). Writing the effective
neutrino magnetic moments in terms of these vectors has one clear
advantage. Maximization (and minimization) of the effective couplings
(with respect to the phases for fixed moduli) can
be understood as a consequence of vector misalignments in parameter
space, thus allowing a quantitative determination of the role played
by the CP-violating phases $\varphi_1$ and $\varphi_3$.

Using Eqs. (\ref{eq:coupling_in_terms_of_vectors}) and
(\ref{eq:vector_definition}) a general analysis, valid for any value
of the $\delta$ Dirac CP phase, can be done. Here instead we fix its
value to $\pi$ as suggested by global fits to neutrino oscillation
data~\footnote{This value differs by about $7\%$ with respect to the
  actual best fit point value for the normal ordering $\delta=1.08\pi$
  \cite{deSalas:2020pgw}.}. In doing so, the entries of the lepton
mixing matrix do not depend anymore on $\delta$ and so we adopt the
notation $U_{\alpha i}\to f_{\alpha i}$. The misalignment as well gets
simplified and involves only the new sources of CP violation. Both the
coefficients and misalignments for this case are shown in
Tab. \ref{tab:flavored_coefficients}. Note that combined analyses of
future DUNE \cite{DUNE:2015lol}, T2HKK \cite{Hyper-Kamiokande:2016srs}
and MOMENT \cite{Cao:2014bea} data will achieve at most a $\sim 10\%$
resolution \cite{Tang:2019wsv}.  Furthermore, from the practical point
of view, such a value facilitates the presentation and analytical
treatment of the various parameter space features we now discuss. It
is worth emphasizing that our results, in particular their qualitative
features, are independent upon our choice.

Let us consider first the parameter space maximization of the
effective couplings, for which \textit{given a data set the more
  stringent bounds (consistent with an upper limit) on the Lagrangian
  couplings are obtained}. From
Eq. (\ref{eq:coupling_in_terms_of_vectors}) it is clear that this
happens whenever the second term acquires its minimum value. For this
to be the case, the flavor vectors, $\vec{\Lambda}_{i\alpha}$, have to
be aligned in certain directions in parameter space. Such
configurations can be determined through the conditions:
\begin{equation}
  \label{eq:minimization_flavored_vectors}
  \frac{\partial}{\partial\varphi_a}
  \left|
    \sum_{i=1}^3 \vec{\Lambda}_{i\alpha}
  \right|^2=0\ ,\quad(a=1,3)\ ,
\end{equation}
that result in relations between the phases and the flavor
vectors. For reactors and one of the delayed components in $\pi$-DAR
($\alpha=e$) we find
\begin{alignat}{3}
  \label{eq:phases}
  \cos(\varphi_1)&=
  \frac{-\Lambda_{1e}^2 - \Lambda_{2e}^2 + \Lambda_{3e}^2}{2\Lambda_{1e}\Lambda_{2e}}\ ,
  \nonumber\\
  \cos(\varphi_3)&=
  \frac{-\Lambda_{1e}^2 + \Lambda_{2e}^2 + \Lambda_{3e}^2}
  {2\Lambda_{2e}\Lambda_{3e}}\ .
\end{alignat}
For the prompt and delayed muon neutrino components in $\pi$-DAR, the
relation follows from Eq. (\ref{eq:phases}) by trading
$\Lambda_{ie}\to \Lambda_{i\mu}$ for $i=1,3$ and
$\Lambda_{2e}\to -\Lambda_{2\mu}$. For the tau neutrino case the same
relations hold. This can be understood as follows. The interference
terms in the muon neutrino case differ from those in the electron
neutrino case by minus signs in the terms involving the
$\Lambda_{2\mu}$ coupling, while in the tau neutrino effective
coupling they differ by a global minus sign. Of course relations in
Eq. (\ref{eq:phases}) along with those for the other flavors should be
combined with the obvious constraint $|\cos(\varphi_a)|\leq 1$, which
in turn fixes the regions in parameter space where minimization of the
flavor vector in (\ref{eq:coupling_in_terms_of_vectors}) is achieved.

We have found that in those parameter space regions, it holds that
$\left|\sum_{i=1}^3 \vec{\Lambda}_{i\alpha} \right|^2_\text{min}=0$.
This means that although the effective neutrino transition magnetic
moment depends upon five parameters (in the CP-violating case), when
moving to the regions where $\mu_{\nu_\alpha}^2$ peaks (namely
$\mu_{\nu_\alpha}^2|_\text{max}$) the mapping of experimental
information (limits or measurement) reduces to a three parameter
problem
\begin{equation}
  \label{eq:mu_nu_alphaSq_max}
  \mu_{\nu_\alpha}^2|_\text{max} = \left|\vec{\Lambda}\right|^2\ ,
\end{equation}
where the parameter reduction follows from the two constraints imposed
by the minimization conditions in
(\ref{eq:minimization_flavored_vectors}). In the $\Lambda_i-\Lambda_j$
plane, the regions obtained this way correspond to the lower boundary
where the couplings (i.e. with smallest possible values) still
saturate an experimental upper limit. As can be seen from
Eq.~(\ref{eq:mu_nu_alphaSq_max}), in this boundary the typical size of
the couplings amounts to that of the experimental upper limit.

\begin{table}
  \renewcommand{\tabcolsep}{0.3cm}
  \renewcommand{\arraystretch}{1.1}
  \centering
  % 1st row
  \begin{tabular}{|c|c|c||c|c|c||c|c|c|}\hline
    \multicolumn{3}{|c||}{\textbf{Electron}}
    & \multicolumn{3}{|c||}{\textbf{Muon}}
    & \multicolumn{3}{|c|}{\textbf{Tau}}\\\hline\hline
    % 2nd row
    $f_{1e}$ & \multicolumn{2}{|c||}{$c_{12}c_{13}$}
    & $f_{1\mu}$ & \multicolumn{2}{|c||}{$c_{23}s_{12} - s_{23}c_{12}s_{13}$}
    & $f_{1\tau}$ & \multicolumn{2}{|c|}{$s_{23}s_{12} + c_{23}c_{12}s_{13}$}\\\hline
    % 3rd row
    $f_{2e}$ & \multicolumn{2}{|c||}{$s_{12}c_{13}$}
    & $f_{2\mu}$ & \multicolumn{2}{|c||}{$c_{23}c_{12} + s_{23}s_{12}s_{13}$}
    & $f_{2\tau}$ & \multicolumn{2}{|c|}{$-s_{23}c_{12} + c_{23}s_{12}s_{13}$}\\\hline
    % 4th row
    $f_{3e}$ & \multicolumn{2}{|c||}{$s_{13}$}
    & $f_{3\mu}$ & \multicolumn{2}{|c||}{$s_{23}c_{13}$}
    & $f_{3\tau}$ & \multicolumn{2}{|c|}{$c_{23}c_{13}$}\\\hline\hline
    % 5th row
    $\widehat{e}_{1e}\cdot \widehat{e}_{2e}$
    & \multicolumn{2}{|c||}{$\cos(\varphi_1)$}
    & $\widehat{e}_{1\mu}\cdot \widehat{e}_{2\mu}$
    & \multicolumn{2}{|c||}{$\cos(\varphi_1+\pi)$}
    & $\widehat{e}_{1\tau}\cdot \widehat{e}_{2\tau}$
    & \multicolumn{2}{|c|}{$\cos(\varphi_1)$}\\\hline
    % 6th row
    $\widehat{e}_{1e}\cdot \widehat{e}_{3e}$
    & \multicolumn{2}{|c||}{$\cos(\varphi_3-\varphi_1+\pi)$}
    & $\widehat{e}_{1\mu}\cdot \widehat{e}_{3\mu}$
    & \multicolumn{2}{|c||}{$\cos(\varphi_3-\varphi_1+\pi)$}
    & $\widehat{e}_{1\tau}\cdot \widehat{e}_{3\tau}$
    & \multicolumn{2}{|c|}{$\cos(\varphi_3-\varphi_1)$}\\\hline
    % 7th row
    $\widehat{e}_{2e}\cdot \widehat{e}_{3e}$
    & \multicolumn{2}{|c||}{$\cos(\varphi_3+\pi)$}
    & $\widehat{e}_{2\mu}\cdot \widehat{e}_{3\mu}$
    & \multicolumn{2}{|c||}{$\cos(\varphi_3)$}
    & $\widehat{e}_{2\tau}\cdot \widehat{e}_{3\tau}$
    & \multicolumn{2}{|c|}{$\cos(\varphi_3)$}\\\hline
  \end{tabular}
  \caption{Coefficients of the flavor-dependent vectors that define the effective
    neutrino magnetic moment coupling in Eq. (\ref{eq:coupling_in_terms_of_vectors}). Shown
    as well are the misalignment products whose values determine the amount of CP violation
    the new couplings come along with. These expression follow from fixing $\delta=\pi$,
    a value that differs from that suggested by neutrino oscillation data by $\sim 7\%$
    (for the normal spectrum case) \cite{deSalas:2020pgw}. The notation 
    $c_{ij}\equiv \cos\theta_{ij}$ and $s_{ij}\equiv \sin\theta_{ij}$, where $\theta_{ij}$
    corresponds to neutrino mixing angles, has been used.}
  \label{tab:flavored_coefficients}
\end{table}

Condition (\ref{eq:minimization_flavored_vectors}) along with the
results in Eq. (\ref{eq:phases}) (and the corresponding ones for the
muon and tau neutrino cases) define the maximum value for the
effective coupling when varying the phases.  On the other hand,
conditions (\ref{eq:minimization_flavored_vectors}) also lead to
critical points that correspond to a maximum, which satisfies:
\begin{equation}
  \label{eq:blind_spot_general}
  \left|\sum_{i=1}^3\vec{\Lambda}_{i\alpha}\right|^2
  =\left|\vec{\Lambda}\right|^2\ .
\end{equation}
When this relation is satisfied, the presence of a neutrino magnetic
transition moment interaction in a particular scattering experiment
will not show up, allowing for large values of fundamental couplings,
even orders of magnitude above the effective coupling limit. We refer
to these regions in parameter space as \emph{blind spots} (identified
for the first time in Ref. \cite{Canas:2016kfy}).  Around these
regions, the smallness of the effective neutrino magnetic transition
moment has little---if nothing---to do with the absolute size of the
$\Lambda_i$ couplings. How large or small $\mu_{\nu_\alpha}^2$
becomes, depends on how much these vectors are aligned in parameter
space. If properly chosen, stringent experimental limits can be
satisfied with couplings that exceed the experimental limit in several
orders of magnitude. This demonstrates that analyses using only
reactor or $\pi-$DAR neutrinos cannot exclude the presence of these
interactions, and hence, both datasets should be included for a full
picture.

The analysis of these regions in parameter space correspond to
$\mu_{\nu_\alpha}^2|_\text{min}$, in contrast to our previous
discussion in which we determined
$\mu_{\nu_\alpha}^2|_\text{max}$. Notice that
  while the latter fixes a lower boundary, the former instead fixes
an upper one. Thus, the largest possible couplings consistent with
experimental information are therefore encompassed within this
region. Of course if the only available information is an upper limit,
smaller values for the couplings are allowed, but those that saturate
the limit are enclosed by this region. Blind spots are found whenever
the following two conditions are simultaneously satisfied
\begin{equation}
  \label{eq:blind_spots_conditions}
  \widehat{e}_{i\alpha}\cdot \widehat{e}_{j\alpha}=1\ ,
  \qquad
  |\Lambda_i|=f_{i\alpha}\left|\vec{\Lambda}\right|\ .
\end{equation}
From Tab. \ref{tab:flavored_coefficients} one can see that
nonvanishing phases are required to access blind spots. For the
electron neutrino effective coupling $\varphi_1=0$ and
$\varphi_3=-\pi$, while for the muon and tau neutrino cases
$\varphi_1=-\pi$ and $\varphi_3=0$. Perturbations of the conditions in
Eq. (\ref{eq:blind_spots_conditions}) such that
Eq.~(\ref{eq:blind_spot_general}) becomes
$|\vec{\Lambda}|^2-\delta|\Lambda|^2$ allow saturating an upper limit
or matching a potential measurement with large $\Lambda_i$ couplings.

So far we have discussed results for reactor and short baseline
accelerator effective couplings in the mass basis. As can be seen from
Eq. (\ref{eq:flavor_reactor_DAR}) along with
Eq. (\ref{eq:lambda_3times3_parametrization}), in the flavor basis
these couplings acquire rather simple forms
\begin{equation}
  \label{eq:reactor_accelerator_flavor}
  \mu_{\nu_\alpha}^2=\sum_{\beta\neq \alpha}|\Lambda_\beta|^2\ ,
\end{equation}
where $\Lambda_\beta=\mu_\beta - i\epsilon_\beta$ and the relation
between these couplings and those in the mass eigenstate basis are
determined by Eq.~(\ref{eq:lambda_flavor_basis}). In this basis, the
effective couplings have no dependence on CP phases. Thus, stringent
bounds on $\mu_{\nu_\alpha}^2$ translate into tight bounds on the
$\Lambda_\alpha$ on which the effective coupling depends upon. In
contrast to the mass basis, the mapping involves just two parameters
and so the regions allowed by a given limit are those determined by
$\sum_{\beta\neq \alpha}|\Lambda_\beta|^2\leq
\mu_{\nu_\alpha}^2|_\text{Exp}$.
However, information on possible CP-violating effects is lost.

\begin{table}[t]
  \renewcommand{\tabcolsep}{0.4cm}
  \centering
  \begin{tabular}{|c|c|c|c|}\hline
    Type & Experiment & Eff. coupling & 90\% CL limit (range)\\\hline\hline
    Reactor & GEMMA \cite{Beda:2012zz} & $\mu_{\nu_e}$ & $2.9\times 10^{-11}$\\\hline
    $\pi$-DAR & LSND \cite{LSND:2001akn} & $\mu_{\nu_\mu}$ & $6.8\times 10^{-10}$\\\hline
    $\pi$-DAR & DONUT \cite{DONUT:2001zvi} & $\mu_{\nu_\tau}$ & $3.9\times 10^{-7}$\\\hline 
    Solar & Borexino \cite{Borexino:2017fbd} & $\mu_{\nu_e}$& $2.8\times 10^{-11}$\\\hline 
    Solar & XENON1T \cite{XENON:2020rca} & $\mu_{\nu_e}$& $[1.4,2.9]\times 10^{-11}$\\\hline 
  \end{tabular}
  \caption{Current most stringent $90\%$ CL limits (or range in the case of XENON1T)
    on effective neutrino magnetic couplings (normalized to the Bohr magneton). The range
    for XENON1T assumes the signal reported in \cite{XENON:2020rca} is due to this new effective
    coupling.}
  \label{tab:current_limits}
\end{table}
All in all, and as we have already stressed, when mapping experimental
upper limits or an actual measurement into parameter space (in the
mass eigenstate basis) the region of interest is the one with
boundaries given by $\mu_{\nu_\alpha}^2|_{\text{max}}$ (lower
boundary) and the perturbed blind spot region (upper
boundary). Following this approach, we proceed with the mapping of
experimental upper limits into parameter space and determine whether
different data sets can be reconciled. For that aim we use the $90\%$
CL upper limits shown in Tab. \ref{tab:current_limits}.

\begin{figure}
  \centering
  \includegraphics[scale=0.5]{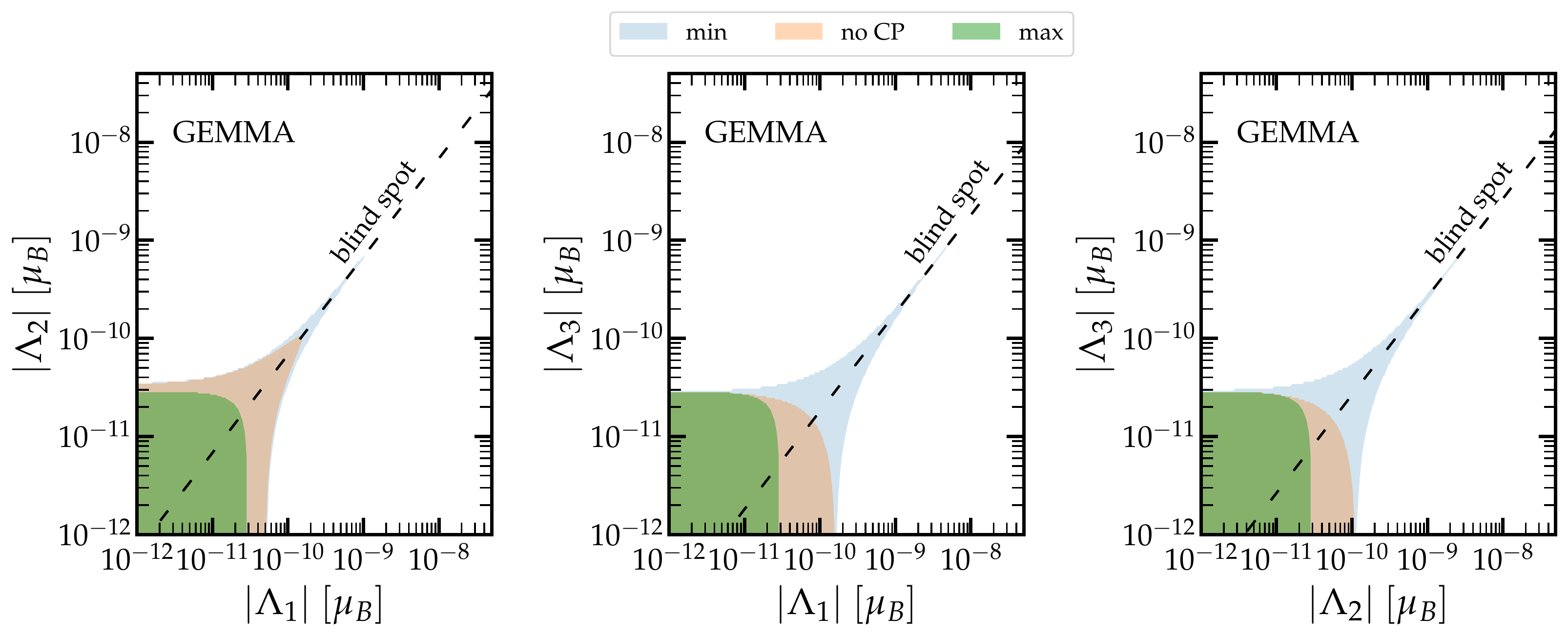}
  \includegraphics[scale=0.5]{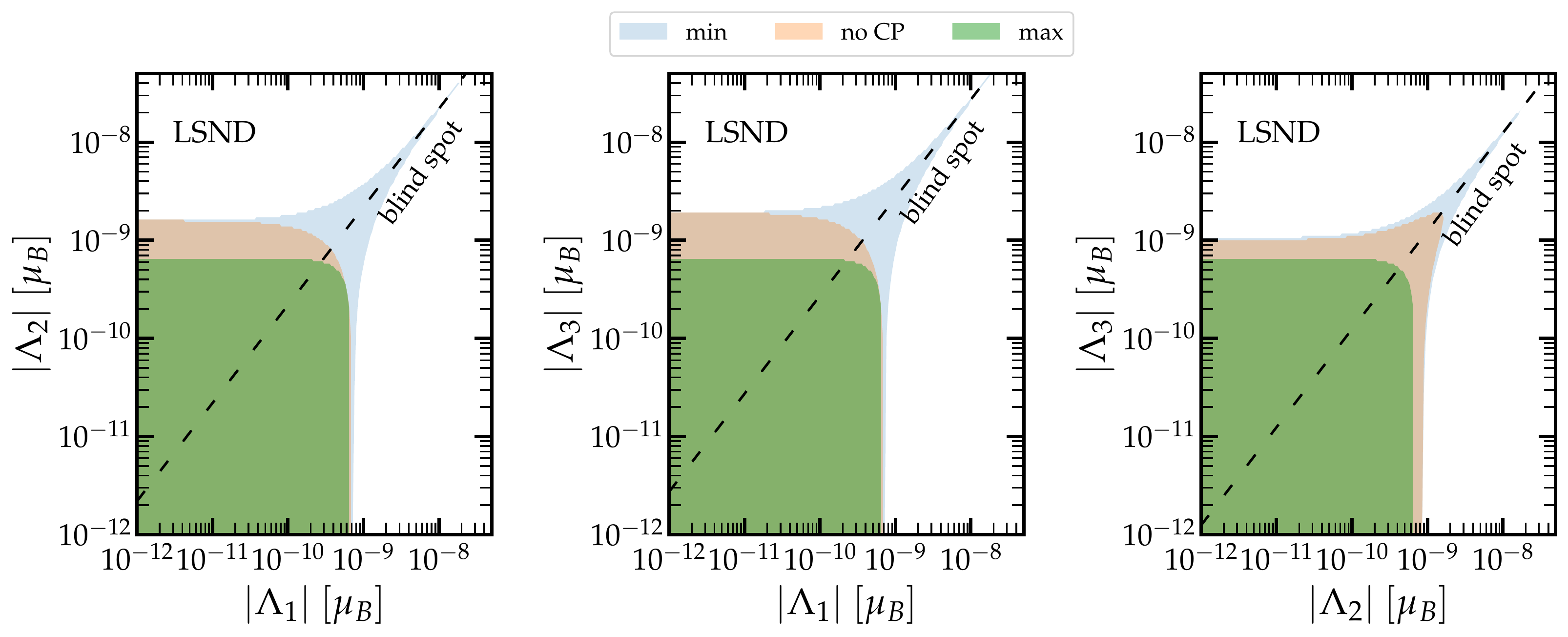}
  \caption{\textbf{Upper row graphs}: $90\%$ CL limits on Majorana
    neutrino transition magnetic and electric dipole moments in Bohr
    magneton units. The results are obtained for couplings in the mass
    eigenstate basis by mapping the upper bound on $\mu_{\nu_e}^2$
    from the GEMMA reactor experiment \cite{Beda:2012zz} to the
    fundamental parameters $\Lambda_i=\mu_i-i\epsilon_i$. Points along
    the colored regions saturate the GEMMA limit while those below
    correspond to smaller values. No CP means the CP phases in the
    electromagnetic sector are fixed to zero. \textbf{Lower row
      graphs}: Same as for graphs in the upper row but for bounds on
    $\mu_{\nu_\mu}^2$ from the LSND accelarator experiment
    \cite{LSND:2001akn}. Results from the DONUT upper limit on
    $\mu_{\nu_\tau}^2$ \cite{DONUT:2001zvi} follow the same
    trend. Since they involve less stringent bounds are not
    displayed.}
  \label{fig:reactor_limits}
\end{figure}
Graphs in the upper row in Fig. \ref{fig:reactor_limits} show the
results obtained by mapping the upper limit on $\mu_{\nu_e}^2$ from
the GEMMA reactor experiment
(Eqs. (\ref{eq:reactor_decat_at_rest_mu_nu}) and
(\ref{eq:coupling_in_terms_of_vectors}) with $\alpha=e$) to the
parameter space of fundamental couplings (those entering in the
Hamiltonian in Eq. (\ref{eq:Lagrangian_M})).  Graphs in the lower row,
instead, follow from limits on $\mu_{\nu_\mu}^2$ from the LSND
experiment. Boundaries in the colored regions saturate the limit,
while points within correspond to smaller values. Shown as well is the
blind spot line along which $\mu_{\nu_\alpha}^2=0$ ($\alpha=e, \mu$).
For parameters along that line, signals in a reactor or a short
baseline accelerator experiment are absent regardless of the size of
the fundamental couplings. Three regions in each $\Lambda_i-\Lambda_j$
plane can be identified. The green region is determined by vector
alignments for which $\mu_{\nu_e}^2$ is maximized, as discussed
previously. For this reason,  in this region the most
constrained parameters are found (the largest couplings have sizes of
the order of the GEMMA or LSND limit). The blue region, in particular
the spike along the blind spot line, follows from parameters that satisfy the alignment condition in
Eq. (\ref{eq:blind_spots_conditions}) but that slightly departure from
the conditions the $\Lambda_i$ should satisfy so to fall exactly in the
blind spot. In this region couplings about two orders of magnitude
larger than the GEMMA or LSND limit can be found, inline with our
previous discussion as well.

To emphasize that the larger the couplings the more aligned the
parameter space vectors $\Lambda_{ie}$ should be, we have as well
included a region where the new interactions have vanishing CP phases
(brown region). One can see that although larger couplings are
allowed, compared with those in the green region, the largest possible
values are not comparable to those in the region where CP phases are
not zero. Finally, it is worth mentioning that the corresponding
results for the $\mu_\tau^2$ coupling follow the same behavior, but
since the upper limit is not as stringent as for the other two
flavors, they are not displayed.

\begin{figure}
  \centering
  \includegraphics[scale=0.5]{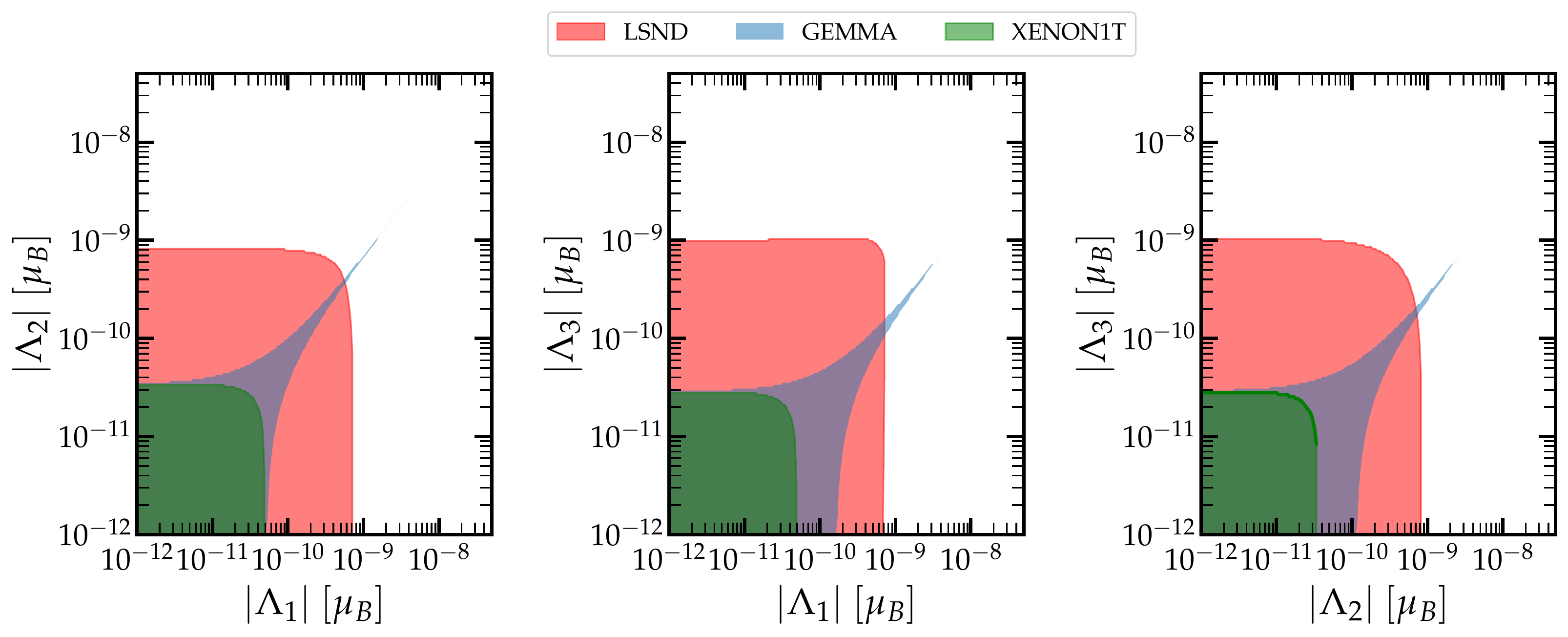}
  \caption{Effective electron and muon neutrino couplings contour
    plots in the planes of fundamental parameters
    $\Lambda_i=\mu_i-i\epsilon_i$ (in Bohr magneton units) in the mass
    eigenstate basis. These regions determine the $90\%$ CL limits on
    the Hamiltonian parameters, they follow from GEMMA, LSND and
    XENON1T measurements
    \cite{Beda:2012zz,LSND:2001akn,XENON:2020rca}. These results
    demonstrate that regions in parameter space where GEMMA and LSND
    limits can be simultaneously satisfied exist.}
  \label{fig:gemma_lsnd_solar}
\end{figure}

Experimental upper limits apply to the effective coupling
$\mu_{\nu_\alpha}^2$ and do not necessarily imply small fundamental
couplings, as demonstrated by the results in
Fig. \ref{fig:reactor_limits}. This then raises the question of
whether limits on the fundamental parameters implied by GEMMA can be
reconciled with those that follow from the less severe LSND
limit. From our discussion of blind spots, it is clear that despite the
large mismatch of these experimental upper limits, regions in
parameter space where this is the case indeed exist. For example, if
couplings and phases are tuned according to the conditions of the
$\mu_{\nu_e}^2$ blind spot, one finds
$\mu_{\nu_\mu}^2=\mu_{\nu_\mu}^2|_\text{max}=|\vec{\Lambda}|^2$
\footnote{We have found that proceeding the other way around,
  \textit{c'est-\`a-dire} fixing couplings according to the blind spot
  conditions for $\mu_{\nu_\mu}^2$, maximizes as well
  $\mu_{\nu_e}^2$.}. Thus, in this region no signal at GEMMA is
expected regardless of the size of the couplings. GEMMA upper bound is
therefore trivially satisfied and couplings with the appropriate size
to satisfy as well the LSND bound are indeed
possible. Fig. \ref{fig:gemma_lsnd_solar} shows the result of a scan
in the parameter space where phases have been fixed such that alignments of the
flavor vectors in $\mu_{\nu_e}^2$ are in place. This choice allows
entering the region where the GEMMA limit can be satisfied with
sufficiently large couplings and hence leads to regions in parameter
space where couplings can simultaneously account for GEMMA and LSND
measurements. 

% ------------------
% Section
% ------------------
\subsection{Effective couplings in solar neutrino experiments}
\label{sec:solar_majorana}
We now turn to the discussion of the effective coupling pertaining
solar neutrinos. In contrast to the reactor and short-baseline cases,
this coupling acquires a rather simple form in the mass eigenstate
basis, as can be seen from Eq. (\ref{eq:mu_nu_solar}). Explicitly it
reads
\begin{equation}
  \label{eq:solar_eff_coupling_mass}
  \mu_{\nu, \text{sol}}^2=\sum_{i=1}^2\left|\Lambda_i\right|^2 
  \left[1 - c_{13}^2P_{ei}^{2\nu}(E_\nu)\right]
  + c_{13}^2\left|\Lambda_3\right|^2\,
\end{equation}
where $P_{ei}^{2\nu}(E_\nu)$ refers to the two-flavor scheme
probability of observing the $i$-th mass eigenstate $\nu_i$ at the
scattering point, given an initial electron neutrino state. No
dependence on CP phases is found in this basis, and so experimental
limits placed on this effective coupling only constrain the mass basis
parameters $\Lambda_i$. The oscillation probabilities introduce as
well an energy dependence. For electron-neutrino scattering
experiments that dependence is to a large degree determined by the
energy range of the solar pp process, while for CE$\nu$NS by the
energy range of the $^8$B reaction. Since these processes peak at
$\sim 0.4\,$MeV and $\sim 10\,$MeV respectively, one can then evaluate
the probability for those energies and then map into parameter space.

\begin{table}
  \renewcommand{\tabcolsep}{0.3cm}
  \renewcommand{\arraystretch}{1.1}
  \centering
  \begin{tabular}{|c|c|c||c|c|c||c|c|c|}
    % 1st row
    \hline
    \multicolumn{3}{|c||}{$\vec{\Lambda}^m_1$}
    & \multicolumn{3}{|c||}{$\vec{\Lambda}^m_2$}
    & \multicolumn{3}{|c|}{$\vec{\Lambda}^m_3$}\\\hline\hline
    % 2nd row
    $\widehat{u}_{1e}\cdot \widehat{u}_{1\mu}$
    & \multicolumn{2}{|c||}{$\cos(\varphi_1+\pi)$}
    & $\widehat{u}_{2e}\cdot \widehat{u}_{2\mu}$
    & \multicolumn{2}{|c||}{$\cos(\varphi_1)$}
    & $\widehat{u}_{3e}\cdot \widehat{u}_{3\mu}$
    & \multicolumn{2}{|c|}{$\cos(\varphi_1+\pi)$}\\\hline
    % 3rd row
    $\widehat{u}_{1e}\cdot \widehat{u}_{1\tau}$
    & \multicolumn{2}{|c||}{$\cos(\varphi_3-\varphi_1)$}
    & $\widehat{u}_{2e}\cdot \widehat{u}_{2\tau}$
    & \multicolumn{2}{|c||}{$\cos(\varphi_3-\varphi_1)$}
    & $\widehat{u}_{3e}\cdot \widehat{u}_{3\tau}$
    & \multicolumn{2}{|c|}{$\cos(\varphi_3-\varphi_1+\pi)$}\\\hline
    % 4th row
    $\widehat{u}_{1\mu}\cdot \widehat{u}_{1\tau}$
    & \multicolumn{2}{|c||}{$\cos(\varphi_3)$}
    & $\widehat{u}_{2\mu}\cdot \widehat{u}_{2\tau}$
    & \multicolumn{2}{|c||}{$\cos(\varphi_3)$}
    & $\widehat{u}_{3\mu}\cdot \widehat{u}_{3\tau}$
    & \multicolumn{2}{|c|}{$\cos(\varphi_3)$}\\\hline
  \end{tabular}
  \caption{Alignments of the parameter space vectors that define the neutrino
    magnetic effective coupling in the solar case in the flavor eigenstate basis
    (see Eqs. (\ref{eq:mu_nu_solar_eff_flavor}) 
    and (\ref{eq:parameter_space_vec_solar_flavor_basis})).
    As in the reactor and short-baseline cases, they ``measure'' the amount
    of CP violation the new physics comes along with.}
  \label{tab:alignments_solar}
\end{table}
In the flavor basis the effective coupling becomes more intricate, but
carries information on possible CP phases. As in the analysis of
reactor and short-baseline experiments, in this case one can as well
understand CP violation in terms of parameter space vectors
misalignments. However,  in contrast to what we have found in those cases,
no alignments in parameter space exist such that blind spots can be
accessed. The coupling in this basis can be written as
\begin{equation}
  \label{eq:mu_nu_solar_eff_flavor}
  \mu_{\nu,\text{sol}}^2 = \sum_{i=1}^3\left|\vec{\Lambda}_i^\text{m}\right|^2
  \left(1 - c_{13}^2 P_{ei}^{2\nu}\right)\qquad
  (\text{with}\,\, P_{e3}^{2\nu}=0)\ .
\end{equation}
Here the oscillation probabilities follow the same meaning as in
Eq. (\ref{eq:solar_eff_coupling_mass}) and the parameter space vectors
are defined according to
\begin{equation}
  \label{eq:parameter_space_vec_solar_flavor_basis}
  \vec{\Lambda}_i^\text{m} = \sum_{\alpha=e,\mu,\tau} \vec{\Lambda}_{i\alpha}^\text{m}
  = \sum_{\alpha=e,\mu,\tau}f_{i\alpha}\Lambda_\alpha\widehat{u}_{i\alpha}\ ,
\end{equation}
where the superindex ``m'' has been introduced to differentiate these
vectors from those in the reactor and short-baseline cases 
given in Eq. (\ref{eq:vector_definition}). Alignments, dictated by the scalar
product of the unit vectors $\widehat{u}_{i\alpha}$, are displayed in
 Tab. \ref{tab:alignments_solar}. We have found that for certain
alignments and parameter choices one of the three vectors
$\vec{\Lambda}^\text{m}_i$ can be set to zero,  however for that particular choice the
other two do not vanish.  Hence, we
conclude that blind spots are not present in the solar neutrino case.

\begin{figure}
  \centering
  \includegraphics[scale=0.5]{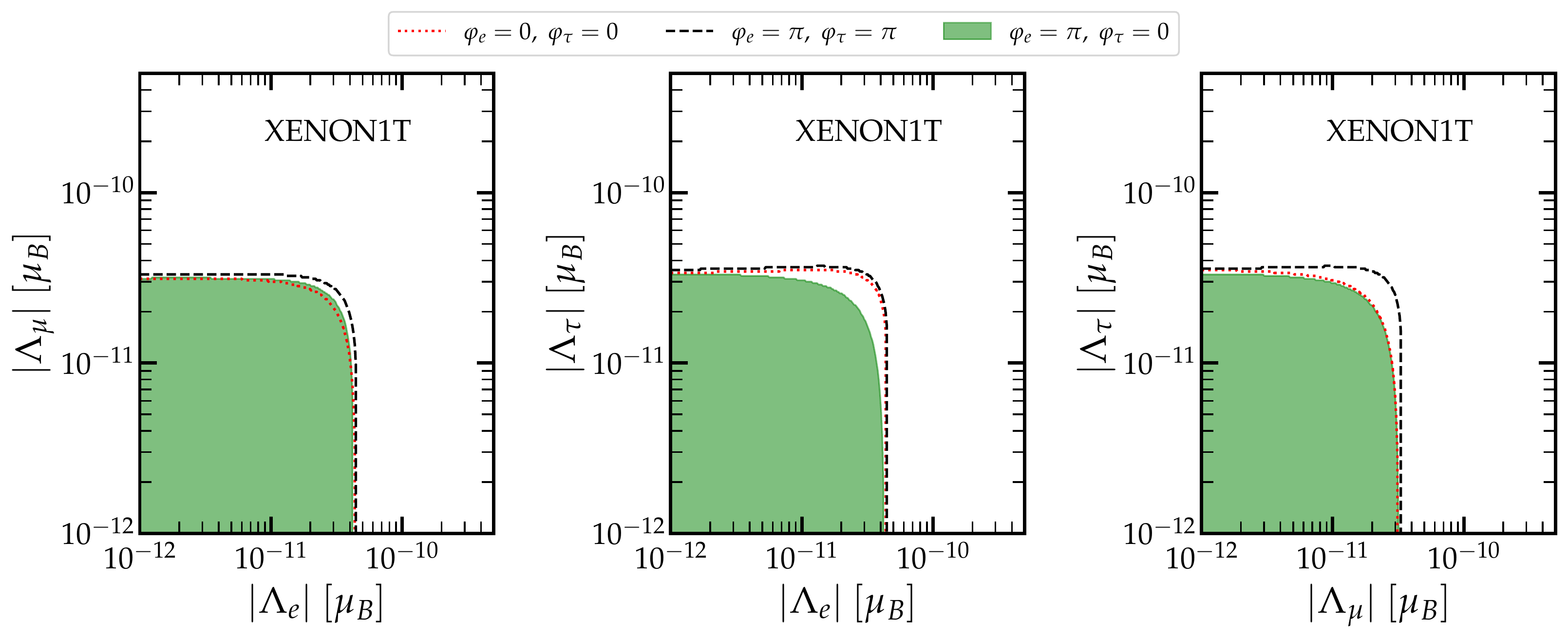}
  \caption{Parameter space regions in the
    $\Lambda_\alpha-\Lambda_\beta$ planes consistent with the $90\%$
    CL range reported by XENON1T (understanding the excess as being
    generated by neutrino magnetic transitions)
    \cite{XENON:2020rca}. Results are obtained for CP phases choices
    such that one of the terms in the effective neutrino magnetic
    moment coupling vanishes. This result demonstrates that even
    though individually one of the terms in $\mu_{\nu_e}^2$ can be
    taken to vanish through proper parameter space vector alignments
    (with suitable CP phases), upper limits on the couplings amount to
    the size of the experimental measurement.}
  \label{fig:solar_xenon_1T_Majorana}
\end{figure}

Results for the parameter space mapping using XENON1T results
(interpreted as being generated by neutrino magnetic transitions) are
shown in Fig. \ref{fig:solar_xenon_1T_Majorana}. The result has been
obtained by fixing the CP phases such that one of the three terms in
Eq. (\ref{eq:mu_nu_solar_eff_flavor}) vanishes. This demonstrates that
although individually one of these terms can be set to zero by proper
alignments (determined by the CP phases) a blind spot region does not
exist. Accordingly,  in view of solar neutrino data the mapping to parameter
space will result always in parameters whose values amount to the same
size of the corresponding measurement.  This in sharp contrast to the
reactor and short-baseline accelerator cases where the blind spot
region allows for couplings with larger values.

% ------------------
% Section
% ------------------
\section{Dirac neutrino magnetic and electric dipole moments}
\label{sec:Dirac_case}
Since the Majorana case allows only for transition moments, in this
case we focus on purely magnetic and electric dipole moments. This
means that in the Dirac coupling matrix in
Eq. (\ref{eq:lambda_3times3_parametrization}) we take the off-diagonal
couplings to be zero. Couplings in the flavor basis follow from the
relations in Eq. (\ref{eq:lambda_flavor_basis}), and so all the
entries of the coupling matrix in that basis are nonzero. In contrast
to the Majorana case, in this particular Dirac scenario all couplings
acquire a rather simple form in both bases. For reactor and
short-baseline accelerator neutrinos in the mass basis one finds
\begin{equation}
  \label{eq:reactor_acc_neutrinos_mass_basis_Dirac}
  \mu_{\nu_\alpha}^2 = \left|\sum_{i=1}^3\vec{\lambda}_{i\alpha}\right|^2\ ,
\end{equation}
where the parameter space vectors in this case read
\begin{equation}
  \label{eq:parameter_space_vec_reac_acc_Dirac}
  \vec{\lambda}_{i\alpha}=f_{i\alpha}\lambda_{ii}\widehat{v}_i\ .
\end{equation}
The unit vectors $\widehat{v}_i$ are orthonormal and the coefficients $f_{i\alpha}$
follow from Tab.~\ref{tab:flavored_coefficients}. Note that in
contrast to the results in the Majorana case, no destructive
interference is found and so no dependence on CP phases. Thus, for
Majorana neutrinos and reactor and short-baseline accelerator
experiments, CP phases play a crucial role but none for purely magnetic
and dipole moments (only possible in the Dirac case). In the flavor
basis, again, expressions for the effective neutrino magnetic moments
are reduced to rather simple form with no CP phases dependence as
well:

\begin{equation}
  \label{eq:flaovr_basis_Dirac_acc_reactor}
\mu_{\nu_\alpha}^2 = \sum_\beta \left|\lambda_{\alpha\beta}\right|^2\ . \qquad  
\end{equation}
For solar neutrinos in the mass basis the result is rather simple and
as expected involves a dependence on the neutrino flavor oscillation
probability
\begin{equation}
  \label{eq:solar_neutrinos_mass_basis_Dirac}
  \mu_{\nu,\text{sol}}^2 = \sum_{i=1}^2 \left|\lambda_{ii}\right|^2\,c_{13}^{2} P_{ei}^{2\nu}
  - \left|\lambda_{33}\right|^2\,s_{13}^2\ ,  \qquad
\end{equation}
where as in the Majorana case, $P_{ei}^{2\nu}$ measures the
probability of detection of the $i$-th neutrino mass
eigenstate. Results for the flavor basis follow directly from
Eq. (\ref{eq:solar_neutrinos_mass_basis_Dirac}) through
\begin{equation}
  \label{eq:lambda_ii_couplings_Dirac}
  \lambda_{ii}=\left(U^T\cdot \lambda_D\cdot U\right)_{ii}\ .
\end{equation}
From these results it becomes clear that each term in
Eq. (\ref{eq:solar_neutrinos_mass_basis_Dirac}) involves CP phases,
but given the structure of the full effective coupling no special
features are found. Alignments might be found so a particular
$|\lambda_{ii}|$ term vanishes, but those alignments will not lead to
blind spots.

\begin{figure}
  \centering
  \includegraphics[scale=0.5]{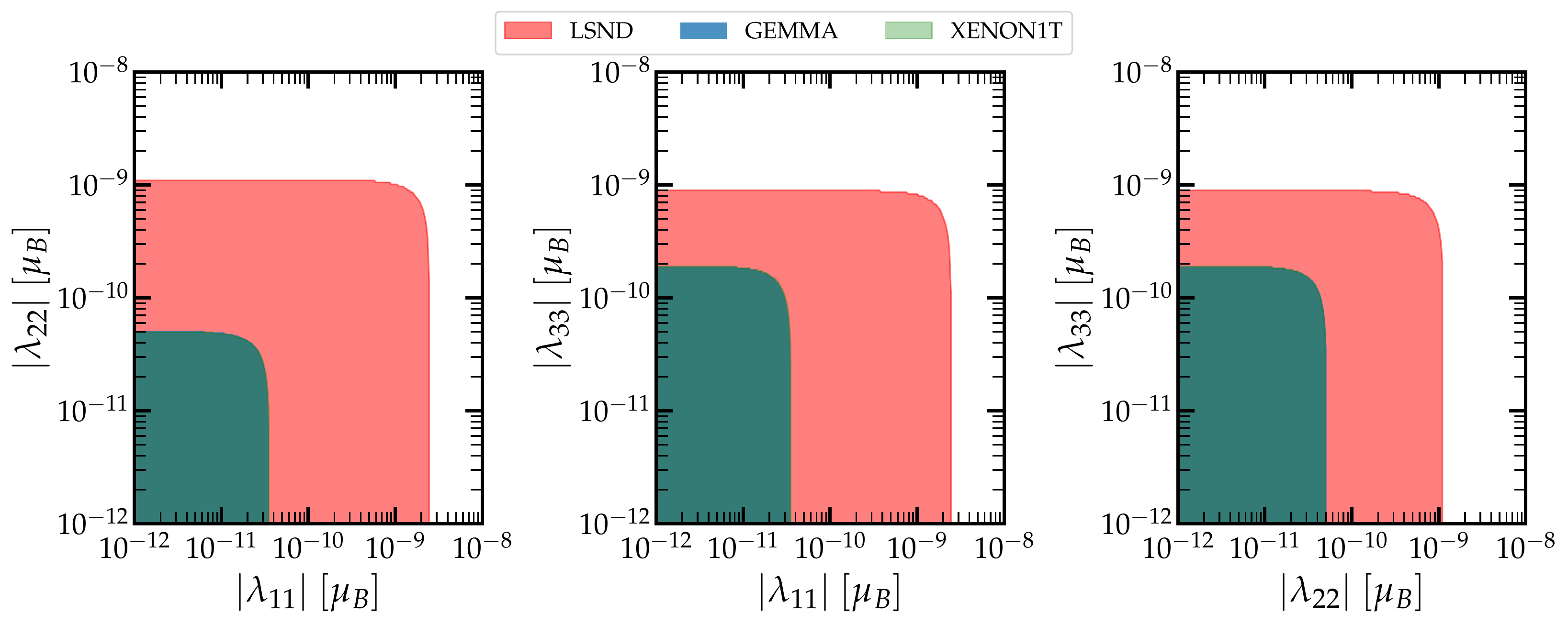}
  \caption{90\% CL limits on the Hamiltonian parameters in the case of
    Dirac neutrino magnetic and electric dipole moments (diagonal
    couplings) in the mass eigenstate basis. Results follow from
    experimental upper limits obtained from neutrino-electron elastic
    scattering processes at LSND \cite{LSND:2001akn}, GEMMA
    \cite{Beda:2012zz} and XENON1T \cite{XENON:2020rca}.}
  \label{fig:Dirac_90CL}
\end{figure}

All in all, in the Dirac diagonal case (pure neutrino magnetic and
electric dipole moments) no features as those found in the Majorana
case (transitions) are observed. As a result, mapping of experimental
data into parameter space implies couplings whose values are of the same
order than the corresponding upper bounds (or an eventual actual
measurement). Results of the mapping are shown in
Fig.~\ref{fig:Dirac_90CL} which displays the 90\% CL limits in the
fundamental parameters derived by considering upper limits obtained
from LSND \cite{LSND:2001akn}, GEMMA \cite{Beda:2012zz}
and XENON1T \cite{XENON:2020rca} neutrino-electron elastic scattering
measurements.

% ---------------------
% Section: Conclusions
% ---------------------
\section{Conclusions}
\label{sec:conclusions}%
We have studied the behavior of effective neutrino magnetic and
electric dipole moments (transitions) and their dependence with the
Hamiltonian (fundamental) parameters, including CP phases in the new
sector. We have considered the case of transitions for Majorana
neutrinos and pure moments for Dirac neutrinos in both the mass and
flavor eigenstate bases. Starting with a generic Dirac CP phase
($\delta$) we have focused on effective couplings valid for reactor
and short-baseline neutrino experiments as well as for solar neutrino
experiments.

We have shown that while individually the flavor effective couplings
are basis dependent, the sum over flavor indices of the three
different couplings is a basis invariant quantity. We thus have
pointed out that if experimental information on the three flavors is
available, the mapping of data to parameter space should be done using
this quantity (see Eq. (\ref{eq:basis_invariant})). Otherwise, when
interpreting data from a given experimental source in terms of
fundamental parameters, the couplings to be used---in the absence of
input on $\delta$---are those we have derived in
Eqs. (\ref{eq:reactor_decat_at_rest_mu_nu}), (\ref{eq:mu_nu_solar}),
(\ref{eq:flavor_reactor_DAR}) and (\ref{eq:flavor_solar}). If instead
one fixes $\delta=\pi$, as suggested by neutrino global fits
\cite{deSalas:2017kay}, a vector treatment of parameter space is
possible and the effective couplings can be readily expressed in terms
of fundamental parameters with the aid of the vectors defined by
Eq. (\ref{eq:vector_definition}) (in the Majorana case) and
Eq. (\ref{eq:parameter_space_vec_reac_acc_Dirac}) (in the Dirac case),
along with the coefficients and alignments defined in
Tabs.~\ref{tab:flavored_coefficients} and
\ref{tab:alignments_solar}. These results provide then an experimental data 
mapping into parameter
space as a straightforward procedure.

We have shown that while short-baseline and reactor couplings are
sensitive to CP phases in the mass basis, the solar neutrino coupling
is instead not. Conversely, in the flavor eigenstate basis
short-baseline and reactor couplings are insensitive to possible CP
phases, while the solar effective coupling acquires a CP-phase
dependence. Though we have found that this behavior is valid
regardless of the neutrino nature, the presence of CP violation does
play a pivotal role in the Majorana case. We have demonstrated that it
enables a blind spot region in parameter space, where the
short-baseline or reactor effective couplings vanish even in the
presence of nonvanishing large fundamental couplings. We, thus, have
pointed out the need for analyses of multiple datasets to remove this
regions in parameter space.

The presence of these blind spots open up regions where stringent
limits, as those implied by GEMMA, can be satisfied with large
Hamiltonian couplings. We have shown that this fact allows reconciling
LSND and GEMMA upper limits which differ in about an order of
magnitude. Here with reconciling we mean finding a common region where
both experimental upper limits can be simultaneously saturated. We
have finally demonstrated that in the Majorana case a tight
short-baseline or reactor upper limit does not necessarily imply
couplings of the order of the experimental bound.
\appendix
\section{Phase average for the solar effective neutrino magnetic
  coupling}
\label{sec:average_solar}
Since in the solar case interference terms come along with phases the
effective parameter $\mu_{\nu_\alpha}$ has to be averaged over
$L/E_\nu$ (the phases involved are $e^{\pm i(\Delta m^2_{ij}/2)(L/E_\nu)}$,
with $i=2,3$ and $j=1$). For the average one can take the normalized
Gaussian smearing function
\begin{equation}
  \label{eq:smearing_Gaussian_function}
  G(L/E_\nu,\langle L/E_\nu\rangle,\sigma)=\frac{1}{\sqrt{2\pi\sigma^2}}\int
  d(L/E_\nu)
  e^{-(L/E_\nu-\langle L/E_\nu\rangle)^2/2/\sigma^2}\ .
\end{equation}
Average over the phases then reads
\begin{equation}
  \label{eq:phases_average}
  \langle e^{\pm i(\Delta m^2_{ij}/2)(L/E_\nu)}\rangle=\frac{1}{\sqrt{2\pi\sigma^2}}\int
  d(L/E_\nu)\,e^{\pm i(\Delta m^2_{ij}/2)(L/E_\nu)}
  e^{-(L/E_\nu-\langle L/E_\nu\rangle)^2/2/\sigma^2}\ .
\end{equation}
This integral can actually be calculated analytically by changing
variables to
\begin{equation}
  \label{eq:new_variables}
  x=\frac{L}{E_\nu}\ ,
  \qquad
  \lambda = \frac{1}{2\sigma^2}\ ,
  \qquad
  x_{\pm}=x - \left(\langle x\rangle \pm i \frac{\Delta m_{ij}^2}{2}\sigma^2\right)\ .
\end{equation}
In terms of these new variables and taking into account the standard
result
\begin{equation}
  \label{eq:standard_integral}
  \int dx\, e^{-\lambda x^2}=\sqrt{\frac{\pi}{\lambda}}\ ,
\end{equation}
the result in Eq. (\ref{eq:phases_averaged}) is obtained
\begin{equation}
  \label{eq:final_result_average}
  \langle e^{\pm i(\Delta m^2_{ij}/2)(L/E_\nu)}\rangle = e^{\pm(\Delta m^2_{ij}/2)\langle L/E_\nu\rangle}
  e^{-\Delta m_{ij}^4\langle L/E_\nu\rangle^2/8}\ .
\end{equation}
\section{Effective couplings in the mass basis in terms of the lepton
  mixing matrix}
\label{sec:majorana_case_leton_mixing_matrix}
In general, one can write the Majorana coupling matrix for the transition
moments as follows (see Eq. (\ref{eq:lambda_3times3_parametrization}))
\begin{equation}
  \label{eq:lambda_coupling_mass_basis}
  \lambda'_{ij} = \epsilon_{ijk} |\Lambda_{k}|e^{i \varphi_k}\ .
\end{equation}
As already noticed, $\varphi_2$ (or any other phase) can be removed by
means of an overall phase scaling of the neutrino fields vector,
leaving behind only two physical CP phases. The notation in
Eq. (\ref{eq:lambda_coupling_mass_basis}) enables writing the
effective coupling in a rather simple form, namely. For a given
neutrino flavor, $\alpha$, in the absence of neutrino flavor
oscillations, the effective neutrino magnetic moment can be rewritten
in terms of the lepton mixing matrix elements according to
\begin{eqnarray}
  \label{eq:eff_nu_mu_lepton_mixing_matrix}
  \mu^2_{\nu_\alpha} &=& \sum_{i,j,l} U_{\alpha l}\,
  \lambda^{\prime\dagger}_{il}\, \lambda^\prime_{ij} \, U^\dagger_{j\alpha}
  \nonumber \\
  &=& \sum_{i,j,l}U_{\alpha l} U^*_{\alpha j}
  \left(
  -\epsilon_{lik}\epsilon_{ijm} |\Lambda_{k}||\Lambda_{m}|e^{-i \varphi_k}e^{i \varphi_m}
  \right) 
  \nonumber \\
  &=& -\sum_{i,j,l}U_{\alpha l} U^*_{\alpha j}
  \left(\delta_{lm}\delta_{kj} - \delta_{lj}\delta_{km}\right)
  |\Lambda_k||\Lambda_m|e^{-i \varphi_k} e^{i \varphi_m}
  \nonumber \\
  &=& \left|\vec{\Lambda}\right|^2 \sum_{l} U_{\alpha l} U^*_{\alpha j}
  - \sum_{j,l} \left(U^*_{\alpha j}|\Lambda_j| e^{-i \varphi_j}\right)
  \left(U_{\alpha l}|\Lambda_l| e^{i \varphi_l}\right)\ .
\end{eqnarray}
After taking into account the unitarity of the lepton mixing matrix
and arranging the second term in
Eq. (\ref{eq:eff_nu_mu_lepton_mixing_matrix}), the effective neutrino
magnetic moment in flavor $\alpha$ becomes
\begin{eqnarray}
\mu^2_{\nu_\alpha} =   \left|\vec{\Lambda}\right|^2 
  - \sum_{j}{|U_{\alpha j}\Lambda_j|}^2 
  % \nonumber \\ 
  - 2\sum_{j > l} \mathbb{R}\text{e}\left[U^*_{\alpha j} U_{\alpha l}
    |\Lambda_j||\Lambda_l| e^{-i\varphi_j} e^{i \varphi_l} \right]\ .
\end{eqnarray}
From this result one can then see that proper alignments determined by
particular choices of the CP phases allow the maximization of the
coupling. More importantly, open the blind spot region discussed in
Sec.~\ref{sec:majorana_case}.
We can see that the previous equation can be written as 
\begin{equation}
\mu^2_{\nu_\alpha} =   \left|\vec{\Lambda}\right|^2 
 -  \left|\left(\sum_{j=1}^3 \left|U_{\alpha j}\right| \left|\Lambda_j\right| \widehat{e}_{j\alpha}\right)\right|^2
\end{equation}
  with $\widehat{e}_{j\alpha}$  unit vectors such that
\begin{equation}
\widehat{e}_{i\alpha} \cdot \widehat{e}_{j\beta}
= \delta_{\alpha\beta}\cos\left[-\arg{\left(U_{\alpha i}\right)} 
+ \arg{\left(U_{\beta j}\right)} - \varphi_i + \varphi_j\right]\ .
\end{equation}
In particular, for the case of $\delta = \pi$, we have the products
$\widehat{e}_{i\alpha} \cdot \widehat{e}_{j\beta}$ given in Table
\ref{tab:flavored_coefficients}.

\section*{Acknowledgments}%
Work supported by CONACYT-Mexico under grant A1-S-23238.  O. G. M.
has been supported by SNI (Sistema Nacional de Investigadores).  The
research of DKP is co-financed by Greece and the European Union
(European Social Fund- ESF) through the Operational Programme ``Human
Resources Development, Education and Lifelong Learning'' in the
context of the project ``Reinforcement of Postdoctoral Researchers -
2nd Cycle'' (MIS-5033021), implemented by the State Scholarships
Foundation (IKY).

\bibliography{references}
\end{document}